\documentclass[conference]{IEEEtran}
\IEEEoverridecommandlockouts
\usepackage{cite}
\usepackage{amsmath,amssymb,amsfonts}
\usepackage{algorithm}
\usepackage{circledsteps}
\usepackage{textcomp}
 \usepackage{array}        
\usepackage{tikz}         
\usepackage{caption}      
\usepackage{mathtools}
\usepackage{algpseudocode}
\usepackage{graphicx}
\usepackage{comment}
\usepackage{url}
\usepackage{textcomp}
\usepackage{multirow}
\usepackage{xcolor}
\usepackage{caption}
\usepackage{subcaption}
\usepackage{enumitem}

\usepackage[normalem]{ulem}

\def\BibTeX{{\rm B\kern-.05em{\sc i\kern-.025em b}\kern-.08em
    T\kern-.1667em\lower.7ex\hbox{E}\kern-.125emX}}

\begin{document}
\title{Consolidating TinyML Lifecycle with Large Language Models: Reality, Illusion, or Opportunity?}

\author{\IEEEauthorblockN{
Guanghan Wu\IEEEauthorrefmark{3},
Sasu Tarkoma\IEEEauthorrefmark{3}, 
Roberto Morabito\IEEEauthorrefmark{1}\IEEEauthorrefmark{3}}

\IEEEauthorblockA{\IEEEauthorrefmark{3}Department of Computer Science,
University of Helsinki, Finland.}
\IEEEauthorblockA{\IEEEauthorrefmark{1}Department of Communication Systems, EURECOM, France.}}

\maketitle
\pagestyle{plain}

\begin{abstract} 
The evolving requirements of Internet of Things (IoT) applications are driving an increasing shift toward bringing intelligence to the edge, enabling real-time insights and decision-making within resource-constrained environments. Tiny Machine Learning (TinyML) has emerged as a key enabler of this evolution, facilitating the deployment of ML models on devices such as microcontrollers and embedded systems. However, the complexity of managing the TinyML lifecycle, including stages such as data processing, model optimization and conversion, and device deployment, presents significant challenges and often requires substantial human intervention. Motivated by these challenges, we began exploring whether Large Language Models (LLMs) could help automate and streamline the TinyML lifecycle. We developed a framework that leverages the natural language processing (NLP) and code generation capabilities of LLMs to reduce development time and lower the barriers to entry for TinyML deployment. Through a case study involving a computer vision classification model, we demonstrate the framework’s ability to automate key stages of the TinyML lifecycle. Our findings suggest that LLM-powered automation holds potential for improving the lifecycle development process and adapting to diverse requirements. However, while this approach shows promise, there remain obstacles and limitations, particularly in achieving fully automated solutions. This paper sheds light on both the challenges and opportunities of integrating LLMs into TinyML workflows, providing insights into the path forward for efficient, AI-assisted embedded system development.
\end{abstract}

\maketitle

\begin{IEEEkeywords}
TinyML, Large Language Models (LLMs), Lifecycle Automation, Embedded IoT Systems,
MLOps for TinyML, Edge AI.
\end{IEEEkeywords}

\pagestyle{plain}

\maketitle

\section{Introduction}

The rapid development of Internet of Things (IoT) technologies and applications has generated a fast-growing need for the development and deployment of machine learning (ML) models at the edge on resource-constrained devices, a paradigm typically called \textit{TinyML} \cite{abadadeComprehensiveSurveyTinyML2023}. TinyML refers to the deployment of machine learning models on resource-constrained devices such as microcontrollers. Unlike traditional ML, which typically runs on powerful cloud or edge servers, TinyML models must operate under tight limitations in memory, compute power, and energy, making their development and lifecycle management significantly more complex \cite{capogrossoMachineLearningOrientedSurvey2024}. TinyML aims to provide intelligent capabilities to small-size factor and power-efficient devices, unlocking ML-powered applications in smart home automation, health, and industrial monitoring. However, the lifecycle management of TinyML models presents unique challenges that demand high human intervention most of the time.

The ML model lifecycle includes various stages, from initial development to deployment and maintenance, typically organized as a pipeline where each component processes data and feeds its output into the next stage. This includes data processing, model training and optimization, and final deployment onto end devices. Managing the TinyML lifecycle on constrained IoT devices, with their limited computing resources and diverse hardware configurations, adds layers of complexity. The landscape becomes further complicated by the heterogeneity of end-device hardware and software, the use of different AI accelerators, and the broad range of device capabilities—all of which make TinyML lifecycle management particularly challenging.

This complex and diverse ecosystem has boosted significant interest in developing automated solutions, which is especially pronounced in large-scale IoT deployments, where hundreds or thousands of edge devices may be deployed across smart cities, industrial networks, or low-power autonomous systems. In these contexts, the efficient and scalable management of ML models becomes essential, as manual intervention is resource-intensive and often impractical.

At the other end of recent AI advancements, Large Language Models (LLMs) have demonstrated remarkable capabilities in natural language processing, code generation, and tasks automation. While LLMs require substantial computational resources and data, their advanced capabilities have inspired research into their potential applications within the IoT context. For example, prior studies have explored the intersection between Generative AI and IoT \cite{sai2024empowering}, envisioning how LLMs can assist developers with code-related tasks, such as creation, completion, and debugging, to support IoT applications development. However, applying LLMs to automate complex lifecycle management tasks for TinyML workflows remains largely unexplored.

Our work takes an initial step toward addressing this gap by leveraging LLMs not only for code generation but also to streamline and automate key TinyML lifecycle components, representing an additional contribution to the intersection of Generative AI and IoT. Specifically, we developed a framework that incorporates the natural language understanding and code generation capabilities of GPT-4o, a leading LLM, as a component within a broader system for automating TinyML lifecycle stages. The framework itself plays a critical role in coordinating these stages within the pipeline, managing interactions with the LLM, and adapting each stage's outputs to ensure compatibility with constrained devices. Unlike traditional TinyML toolchains (e.g., TensorFlow Lite \cite{davidTensorFlowLiteMicro}, Edge Impulse \cite{hymelEdgeImpulseMLOps2023}) that require specialized human involvement across multiple steps, our framework focuses on minimizing developer effort by using LLMs to automate key tasks such as configuration generation, conversion scripting, and deployment sketch creation. Rather than replacing these toolchains, our framework is designed to be complementary, offering an automation layer particularly beneficial in hyperscale IoT deployments, where managing tailored workflows across numerous heterogeneous devices can become increasingly cumbersome.

\begin{figure*}[!t]
    \centering
    \includegraphics[width=0.7\textwidth]{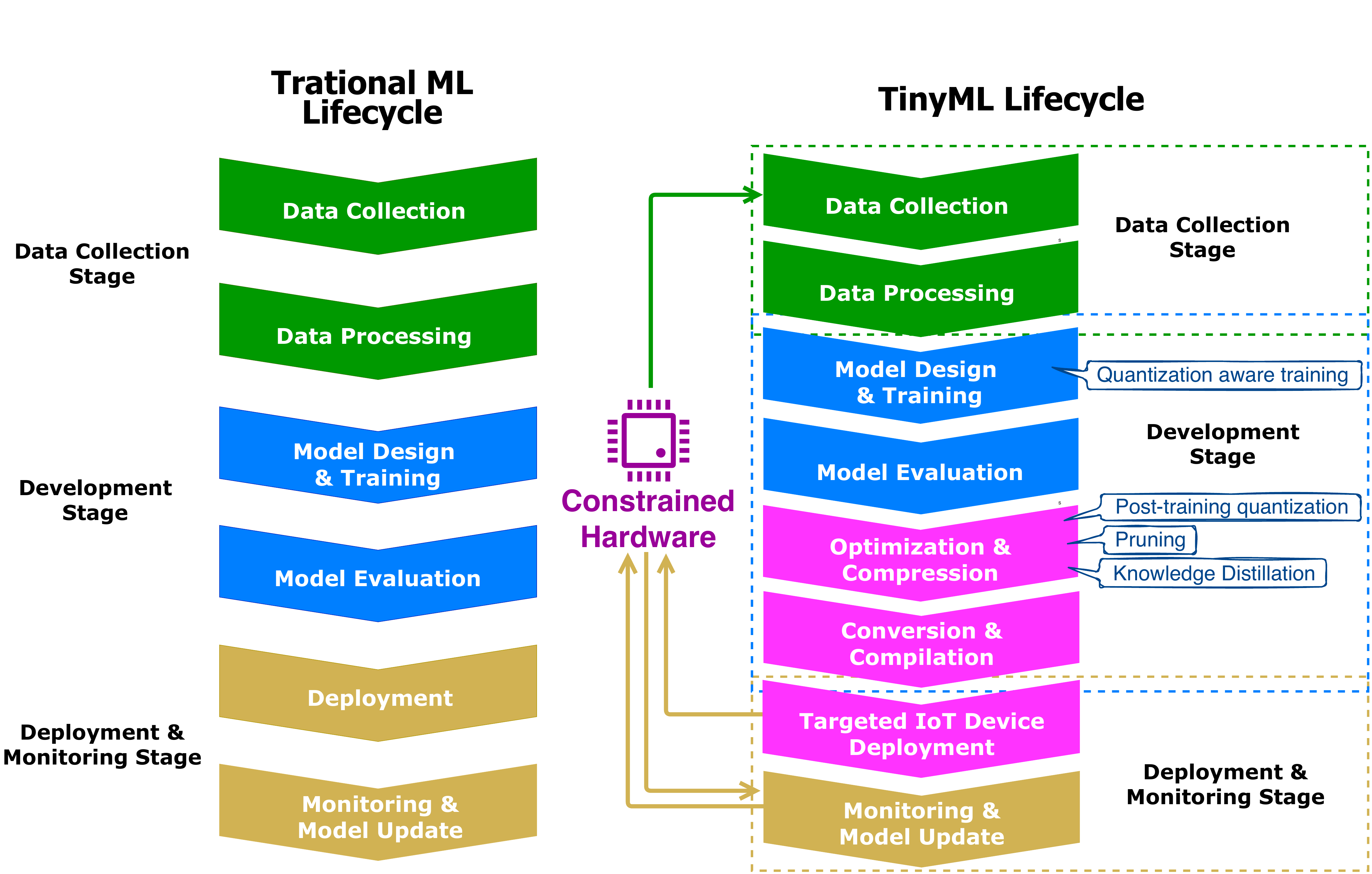}
    \caption{Comparison of Traditional ML and TinyML Lifecycles. The TinyML lifecycle introduces additional steps such as model optimization, compression, and targeted IoT deployment to adapt to constrained hardware environments. Techniques like quantization-aware training, pruning, and knowledge distillation are employed to ensure efficient model performance on resource-limited devices, highlighting the unique adaptations required for TinyML applications.}
    \label{fig:lifecycle}
\end{figure*}

In this paper, we introduce the main requirements and components of our framework and present a detailed case study involving a deep neural network-based classification application on a typical constrained IoT device to evaluate the effectiveness of our approach. Our empirical findings demonstrate that LLM-powered automation can enhance the efficiency and adaptability of TinyML development, offering a promising approach for innovation in AI-assisted IoT systems. At the same time, we recognize that current LLM technologies and our framework have notable limitations, highlighting that full automation may still remain somewhat aspirational.  Based on our experience, we also address the challenges and opportunities identified, outlining a roadmap for future exploration.

\section{Navigating the TinyML Lifecycle}

With the extension of ML models to IoT devices, TinyML systems face several constraints that present three key challenges:
\\
\Circled{1} \underline{Computational Resource Constraints:} Edge devices, particularly microcontrollers, necessitate specialized model architectures to function within their limited processing power and memory capacities. TinyML target devices, such as microcontrollers, typically have kilobyte-scale memory, reduced computational unit sizes, low clock frequencies, and simplified architectural features, often supporting only integer operations \cite{capogrossoMachineLearningOrientedSurvey2024}. For instance, the Arduino Nano 33 BLE Sense, a widely used TinyML board, operates with a 64 MHz clock, 1 MB of flash memory, and 256 KB of RAM, exemplifying these constraints.
\\
\Circled{2} \underline{Efficiency and Performance Trade-offs:} TinyML models must balance energy efficiency, processing speed, and compact model size against potential reductions in accuracy due to limited resources. For long autonomous operation on low-capacity batteries, TinyML models prioritize efficiency metrics over raw performance. Techniques like model compression and optimization are applied to reduce energy consumption and latency, but these often come at the cost of model precision, requiring developers to carefully weigh performance trade-offs.
\\
\Circled{3} \underline{Platform Diversity and Compatibility:} The significant diversity in computing units and instruction set architectures creates substantial challenges for TinyML development \cite{davidTensorFlowLiteMicro}. This hardware heterogeneity includes variations in specifications, development methodologies, and proprietary libraries, complicating the scalability and compatibility of TinyML applications across platforms \cite{hymelEdgeImpulseMLOps2023}. Additionally, differing user requirements, such as resource and energy constraints, often necessitate the development of device-specific models, underscoring the need for adaptable and compatible solutions.

These challenges highlight the need to rethink various stages of the TinyML lifecycle to enable the effective deployment of ML models on resource-constrained devices like microcontrollers \cite{abadadeComprehensiveSurveyTinyML2023}. Fig. \ref{fig:lifecycle} compares the traditional ML lifecycle with a generic TinyML lifecycle. Both lifecycles consist of three stages: data collection, development, and deployment \& monitoring. However, the TinyML lifecycle involves additional steps and distinctions. For example, TinyML data is typically collected via edge device peripherals, such as sensors, from real or simulated production environments. During the ML model development stage, techniques such as quantization and model compression are applied to address the challenges posed by hardware heterogeneity and constraints specific to each deployment environment. Before deployment, models must be converted to formats compatible with TinyML frameworks (e.g., \texttt{tflite} for TensorFlow Lite) and compiled into forms that can run on microcontrollers, often through the use of dedicated ML software libraries suitable for embedded ML, requiring C/C++ code or a lightweight runtime to interpret the model on the target device \cite{warden2019tinyml}.

To sum up, TinyML development demands interdisciplinary expertise spanning software, hardware, and ML. The field remains nascent, with limited benchmarks and immature tooling, which together present substantial obstacles to widespread TinyML implementation and optimization \cite{hymelEdgeImpulseMLOps2023}. Second, the complexity of seamlessly managing each step of the TinyML lifecycle underscores the need for streamlined and automated mechanisms. This is where we envision that LLMs can play a transformative role.

\section{The Rise of Language Models}

Built upon transformer architectures with self-attention mechanisms, LLMs owe their versatility across various tasks—including text generation, translation, and software development—to their enormous scale and the diversity of data on which they are trained. For example, OpenAI's GPT-4o model, which boasts approximately 1.8 trillion parameters across 120 layers, draws on extensive training datasets spanning diverse domains, allowing it to perform complex tasks and adapt to a wide range of applications \cite{kaddourChallengesApplicationsLarge2023}. 
\\
\\
\textbf{LLMs in Specialized Domains: Software Engineering and Code Generation.} These models have shown transformative capabilities in software engineering, particularly in code generation and understanding. They can autonomously perform tasks ranging from code completion to complex software architecture design. Advanced models like GPT-4o and Codex have demonstrated proficiency across multiple programming languages, often matching human-level performance in various coding challenges \cite{vaithilingamExpectationVsExperience2022}. Beyond generating simple code snippets, LLMs can handle sophisticated programming tasks such as algorithm implementation, API integration, and even translating code across languages, making them highly versatile for developers. LLMs are also capable of understanding code structure and semantics, enabling tasks such as documentation generation, automated debugging and supporting the generation of customized code suited to diverse development environments. For instance, they can optimize generated code to align with specific system constraints, aiding in adaptation for heterogeneous computing environments \cite{seo_flexible_2023}. This adaptability makes LLMs valuable tools for efficient, context-sensitive code generation, aligning well with the resource-constrained and specialized requirements of TinyML applications that we investigate in this work.
\\
\\
\textbf{The Role of Prompt Engineering.} To fully unlock their potential, LLMs often rely on prompt engineering, which is a technique designed to elicit precise and reliable outputs from these models. Crafting effective prompts enables LLMs to follow structured instructions, producing desired outcomes across a range of complex tasks. Effective prompts typically include \textit{role definition}, \textit{instructions}, \textit{context}, \textit{input data}, and \textit{output indicator}. Core prompt engineering techniques, such as Few-Shot Prompting, Chain-of-Thought reasoning, and Self-Consistency prompting, enhance the accuracy and control of LLM responses, especially for multi-step processes \cite{yaoReActSynergizingReasoning2023}. These approaches help to improve LLM performance in tasks like code generation but also address domain-specific challenges by providing clear, step-by-step guidance.
\\
\\
\textbf{LLMs and IoT: An Expanding Intersection.} Generative AI is increasingly shaping advancements in IoT and edge computing, bringing new capabilities to data processing, interaction, sensing, and security \cite{wen2024generative}. Looking specifically at LLMs, as a subarea of GenAI technologies, research reveals several promising directions in this expanding intersection. Expectedly, LLMs are instrumental in \textit{code generation and customization} for IoT, enabling model adaptation across diverse environments, such as federated learning, where models adjust to client-specific data and hardware \cite{seo_flexible_2023}. In the same area, it is demonstrated that advanced models like GPT-4o can generate code for complex embedded systems tasks, such as register-level drivers and power optimization techniques \cite{englhardt2024exploring}. In \textit{personalized device interaction}, fine-tuned LLMs allow IoT devices to respond to natural language commands, making technology more accessible by allowing users to configure devices through conversational interfaces \cite{choaib2024iot}. LLMs also show potential in \textit{sensor data analysis}, supporting real-time interpretation for specialized tasks like gesture recognition, enhancing applications in health monitoring and smart wearables \cite{sooriya2023poster}. In \textit{proactive network security}, LLMs are advancing IoT security through automated vulnerability testing, enhancing protocol fuzzing by extracting protocol information and reasoning about device responses \cite{wang2024llmif}. 
All these applications showcase the versatility of LLMs within IoT. With this work, we open up a new area of exploration on how advanced AI, like LLMs, can enhance embedded ML.
\\
\section{LLMs Requirements Towards Automating the TinyML Lifecycle.} Automating the TinyML lifecycle requires a holistic approach to leveraging LLMs that goes beyond simply generating code. To address the specific constraints and complexity of TinyML applications, an effective framework must incorporate a range of tools, prompt engineering techniques, and adaptive strategies that enable LLMs to interact seamlessly with resource-constrained IoT environments. These requirements can be grouped into key areas:
\begin{itemize}
    \item \textit{Adaptive Prompt Engineering:} Prompt engineering is critical for eliciting reliable, task-specific outputs from LLMs, particularly when managing the unique requirements of TinyML tasks. Different prompt engineering techniques allow the model to better understand and execute multi-step processes. Furthermore, customized prompt templates tailored to different stages of the TinyML lifecycle, such as model quantization or sketch generation, streamline interactions and reduce resource consumption by guiding the LLM with precise, context-specific instructions, ensuring that outputs align with both application goals and device limitations. Furthermore, as the TinyML lifecycle involves multiple inter-dependent stages, LLMs need to ensure seamless consistency among the different lifecycle pipeline stages.
    \item \textit{Error Handling and Iterative Refinement:} Due to the often complex and context-dependent nature of TinyML tasks, a robust framework must include mechanisms for detecting and correcting errors. Iterative prompting and self-consistency checks help the LLM refine its outputs, particularly in stages where minor mistakes could compound into larger issues, such as model deployment or hardware-specific adjustments. These iterative methods help to improve output reliability and to reduce the need for human intervention by enabling the system to self-correct when feasible.
    \item \textit{Tool Integration and Orchestration:} The TinyML lifecycle requires close integration of the LLM with specialized embedded ML libraries, device-specific compilers, and lightweight runtimes. Effective orchestration between these components and the LLM is essential for managing workflows, coordinating data inputs, and ensuring that each output is compatible with the constrained target environment. Middleware LLMs frameworks, such as LangChain for prompt management and LangSmith for monitoring LLM interactions, play a pivotal role in enabling these complex, multi-component interactions.
    \item \textit{Human-in-the-loop Interventions:} Although the goal is to reduce human involvement, certain stages in the TinyML lifecycle may still benefit from expert oversight. A well-designed framework can provide targeted suggestions or alerts for human review when complex or high-stakes tasks are encountered. This allows for a balanced approach, where automation handles routine processes, but human expertise is available for critical interventions.
\end{itemize}

In summary, while LLMs offer powerful capabilities, realizing their full potential for TinyML lifecycle automation requires a carefully designed ecosystem of tools and strategies. In the next section, we introduce how the developed framework travels in the direction of offering a practical solution for AI-assisted TinyML development.

\section{Blueprint for Automation}

\begin{figure}[!t]
    \centering
    \includegraphics[width=\columnwidth]{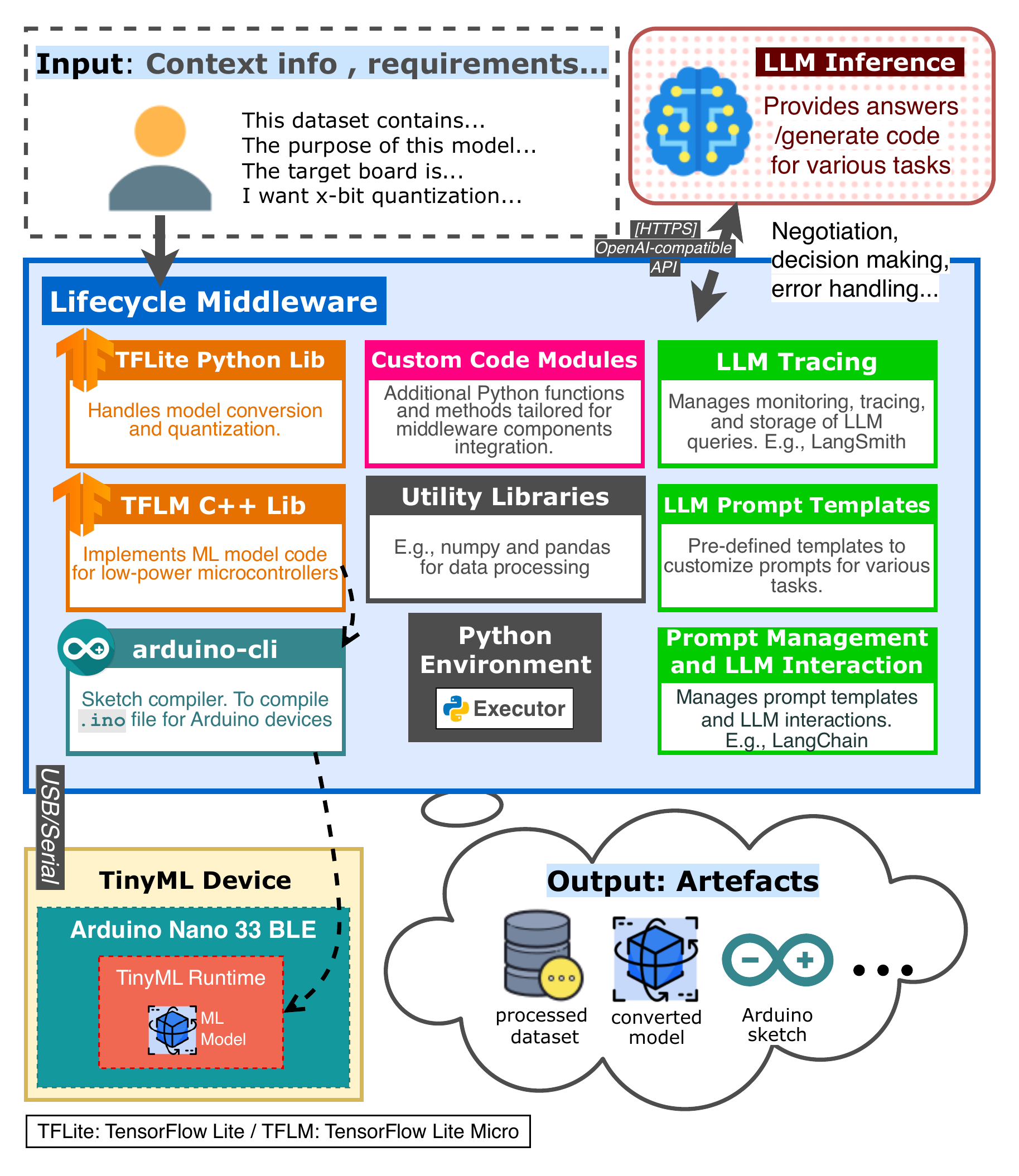}
    \caption{A detailed view of the Lifecycle Middleware components, organized by function. ML Software Libraries (orange) handle model preparation, LLM Integration Components (green) manage interaction with OpenAI's GPT models, Custom Code Modules (dark pink) provide additional framework-specific logic and user interaction, and Utility Libraries (dark gray) support data handling.}
    \label{fig:schema}
\end{figure}

Following the requirements outlined in the previous sections, we introduce our framework designed to integrate OpenAI's GPT family of LLMs with TinyML tools, aimed at automating key stages of the TinyML lifecycle. We present the framework from two perspectives regarding the what and the how, emphasizing both its structural components and operational workflow.

\begin{figure*}[!t]
    \centering
    \includegraphics[width=\textwidth]{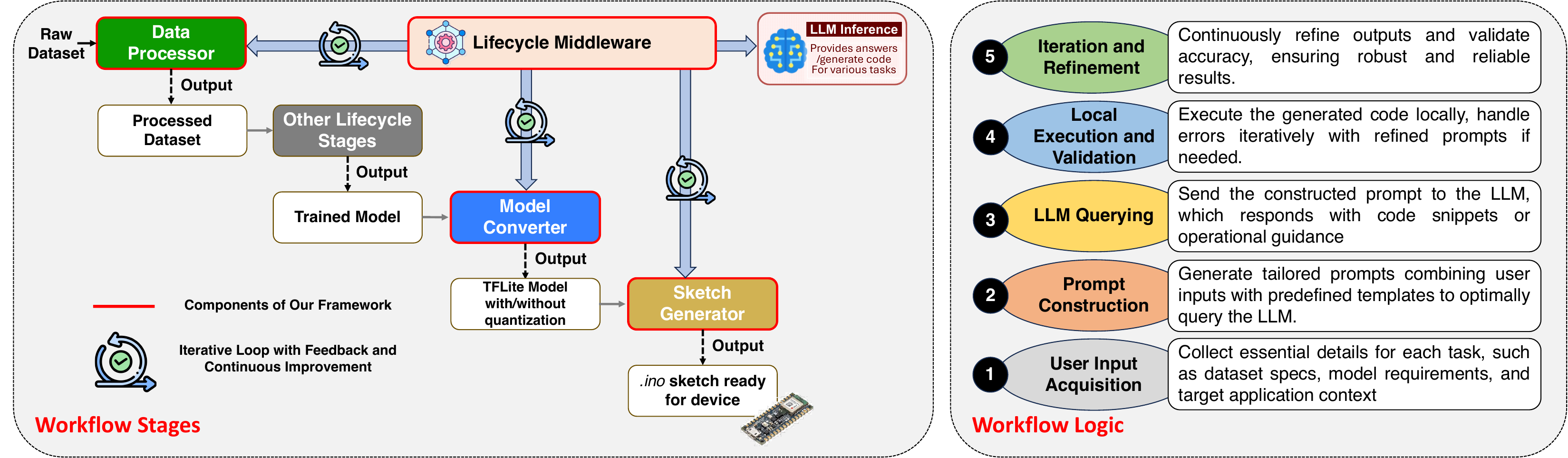}
    \caption{The left panel (\textit{Worfklow Stages}) shows the core functional stages, from data processing to model conversion and sketch generation, guided by a middleware that interacts with the LLM for task-specific code generation and iterative feedback. The right panel (\textit{Workflow Logic}) outlines the five-step logical flow: from acquiring user inputs to iterating for output refinement, offering a detailed overview of the procedural steps involved in TinyML application lifecycle management.}
    \label{fig:implstages}
\end{figure*}

\textbf{The "what" — Framework Components.} As shown in Fig. \ref{fig:schema}, there are several essential building blocks  that shape the ecosystem around our framework. The framework acts as an orchestrator, connecting human-in-the-loop input, LLM interactions, embedded ML libraries, and the hardware/software requirements of target devices. Each component contributes to this ecosystem, with dedicated roles such as model conversion, quantization, dynamic prompt generation, and error handling. Together, these elements enable operational artifacts, robust feedback loops, and adaptable workflows, establishing the framework's technical foundation.

\textbf{The "how" — Process Workflow.} Illustrated in Fig. \ref{fig:implstages}, it focuses on the structured workflow that drives TinyML applications from user input to deployment. The Workflow Stages (left panel) outline the main functional steps: Data Processing, Model Conversion, and Sketch Generation, among others. These stages are orchestrated by the Lifecycle Middleware (detailed in Fig. \ref{fig:schema}) that bridges user-defined goals with LLM-powered code generation. Meanwhile, the Workflow Logic (right panel) presents a five-step logical flow from user input acquisition to iterative refinement, ensuring each stage receives feedback for continuous improvement and precision.

At the heart of the Fig. \ref{fig:schema} framework is the \textit{Lifecycle Middleware}, which integrates multiple specialized components. Custom code modules implement the core logic of the framework and facilitate user interactions, while ML software libraries prepare machine learning models for deployment by managing tasks such as conversion and quantization. Dedicated components for LLM integration coordinate interactions with OpenAI's GPT models, and utility libraries handle essential data processing and manipulation tasks to ensure smooth workflow transitions. Importantly, once the user provides the dataset, the subsequent workflow is fully automated. The framework leverages a set of curated prompt templates, each corresponding to a different lifecycle stage (e.g., data processing, model conversion, deployment). These templates are dynamically adapted based on the task context and feedback signals, such as errors captured during execution, allowing the system to invoke the LLM autonomously and progress through the pipeline without requiring further human intervention.

\begin{figure*}[!t]
    \centering
    \includegraphics[width=1\textwidth]{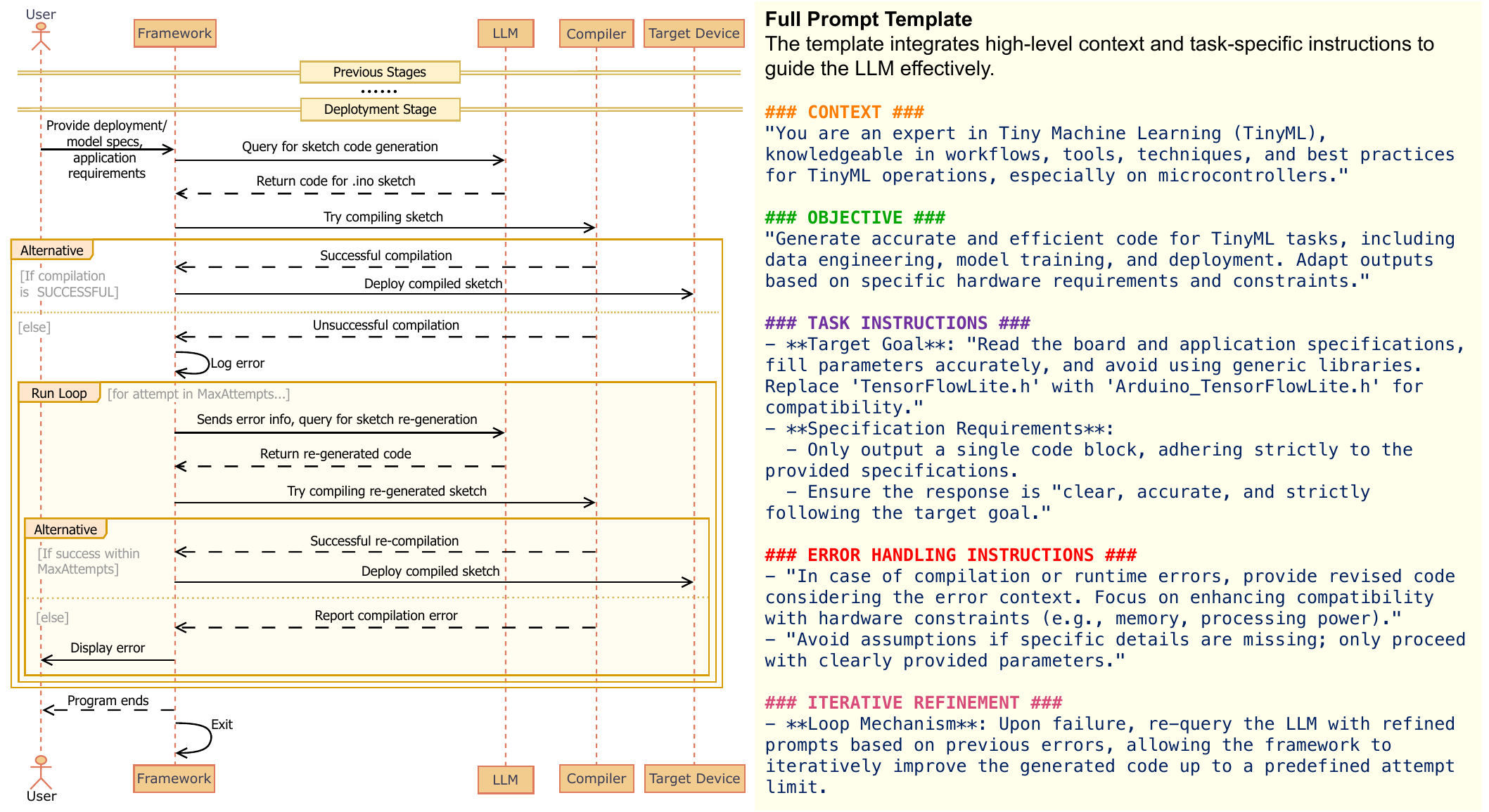}
    \caption{Sequence of the Deployment Sketch Generation. The diagram illustrates the iterative process of generating and deploying an Arduino sketch, showing interactions between the user, framework, LLM, \texttt{arduino-cli} compiler, and target device, with built-in error handling and retry mechanisms. The prompt snippet on the right highlights the structured setup used to guide the LLM in generating code for TinyML tasks, including context setup, task-specific goals, and error-handling protocols.}
    \label{fig:sequence}
\end{figure*}

The \textit{Custom Code Modules} act as the central orchestrator, guiding the flow of user-defined goals such as requirements, preferences, and specifications into the LLM environment. It facilitates seamless communication between the LLM and the framework, ensuring that outputs are continuously refined and adapted to align with the hardware and software constraints of TinyML devices. These modules enable the framework to generate outputs that are then processed by specialized \textit{embedded ML libraries}, such as TFLite, for model conversion and quantization, ensuring alignment with the hardware and software constraints of the target devices. This is achieved by coordinating tasks and managing the iterative exchange with the LLM.

The LLM-related components, depicted in green in Fig. \ref{fig:schema}, handle the interactions with OpenAI's GPT models and play a crucial role in refining outputs iteratively. \textit{LLM Prompt Templates} provide pre-defined structures that guide the LLM in generating context-specific responses, tailored to meet TinyML lifecycle needs. The \textit{Prompt Management and LLM Interaction} module, facilitated by tools like LangChain, manages the flow of prompts and responses, ensuring that each step aligns with the user-defined goals and device constraints. Additionally, \textit{LLM Tracing} tools, such as LangSmith, enable monitoring, tracing, and storage management of LLM interactions, allowing for debugging and optimization throughout the workflow.

Ultimately, the framework generates essential artifacts, including processed datasets, optimized models, and executable sketches that are fully compatible with TinyML devices. 

In each lifecycle stage automated by the framework (Fig. \ref{fig:implstages} left panel), the workflows follow a consistent logic, though each stage, such as data processing, model conversion, or sketch generation, has specific operations and techniques. By adhering to a standardized workflow (Fig. \ref{fig:implstages} right panel), each stage begins by gathering stage-specific information. For example, in data processing, the system collects information about the dataset and model purpose, while in model conversion, it gathers details like dataset overview and quantization requirements.
Using the gathered input, the framework constructs tailored prompts specific to each stage. These prompts are then passed to the LLM to generate configuration settings or code snippets, which the framework subsequently executes or compiles locally. For instance, the system may execute Python code or compile an Arduino sketch, depending on the stage requirements.
To illustrate, in the \textit{data processing stage}, GPT-4o assists in automating preprocessing steps, including data cleaning, normalization, and augmentation, ensuring that the data is prepared for model training with minimal manual intervention. For \textit{model conversion} (optimization and quantization), the framework provides scripts and configurations to convert models into formats suitable for deployment on resource-constrained devices, such as TensorFlow Lite. Finally, in the deployment stage (\textit{sketch generation}), the LLM generates scripts tailored to the specific hardware and application requirements, including code for deploying models on devices like the Arduino Nano 33 BLE. 
It is worth mentioning that our work focuses specifically on representative stages of the ML lifecycle that are unique to TinyML workflows due to the computing constraints and requirements of resource-constrained devices. These include stages like model optimization and deployment, which necessitate additional operations compared to a traditional ML lifecycle. We deliberately omit other stages, such as model training, as these remain largely identical in both TinyML and traditional ML workflows.
For each lifecycle stage, if an error arises during local execution, the system triggers an iterative retry mechanism that engages the LLM with specialized error-handling prompts. This iterative process continues, refining the output until successful execution is achieved or until the maximum retry threshold is reached. The process concludes either with the generation of final artifacts, such as processed datasets, converted models, or Arduino sketches, or with termination if the iteration limit is exceeded.

Fig. \ref{fig:sequence} shows the interactions between the user, framework, LLM, device compiler, and target device during the sketch deployment stage, including the retry mechanisms that ensure robust code generation and deployment. It is worth emphasizing that the framework includes a feedback-driven retry mechanism: when an operation fails (e.g., due to compilation or deployment errors, the system parses the error and updates the prompt accordingly. This process is handled automatically, enabling the LLM to regenerate code that better aligns with the hardware constraints or resolves the encountered issue—without manual prompt rewriting. Instead of displaying the entire prompt code, the snippet on the right provides a structured template summarizing the core components of the prompt, including context, objectives, task-specific instructions, and error handling. This template guides the LLM in generating code that aligns with the hardware constraints and application requirements.

\section{Case Study: From Fruit to Functionality}

When evaluating the implemented system, we aimed to address three key questions (Q\#):

\begin{itemize}
    \item[\textbf{(Q1)}] \textit{\textbf{TASK COMPLETION SUCCESS:} How reliably does the framework and the LLM successfully execute tasks in each developed lifecycle stage?}
    \item[\textbf{(Q2)}] \textit{\textbf{EXECUTION TIME:} What is the time required to complete each stage of the lifecycle, and how does this reflect the framework's efficiency?}
    \item[\textbf{(Q3)}] \textit{\textbf{OPERATIONAL COST:} What are the monetary costs associated with recursive LLM API calls, and how do they balance against the benefits of automation?}
\end{itemize}

These dimensions are inherently interconnected but provide different perspectives on the system's performance. Success rate evaluates the system’s robustness and reliability, time consumption measures efficiency, and monetary cost assesses the practicality of the approach in resource-constrained environments.

To explore these questions, we conducted a detailed case study involving a fruit classification task—representative of computer vision workloads—on the Arduino Nano 33 BLE board. This use case was chosen because it aligns with an officially supported Arduino example, allowing us to validate our LLM-generated sketch against a reliable reference implementation. While simple in nature, it includes the typical lifecycle steps (data processing, quantization, and hardware-specific code generation), and our framework is readily extensible to more complex datasets or embedded ML applications.
From the ML model perspective, we employed a simple computer vision approach using a convolutional neural network (CNN) model for fruit classification. This model is designed to classify different types of fruits (e.g., apples, bananas, oranges) based on their color variations, leveraging the device's built-in RGB color sensor.

\begin{figure}[!t]
    \centering
    \includegraphics[width=\columnwidth]{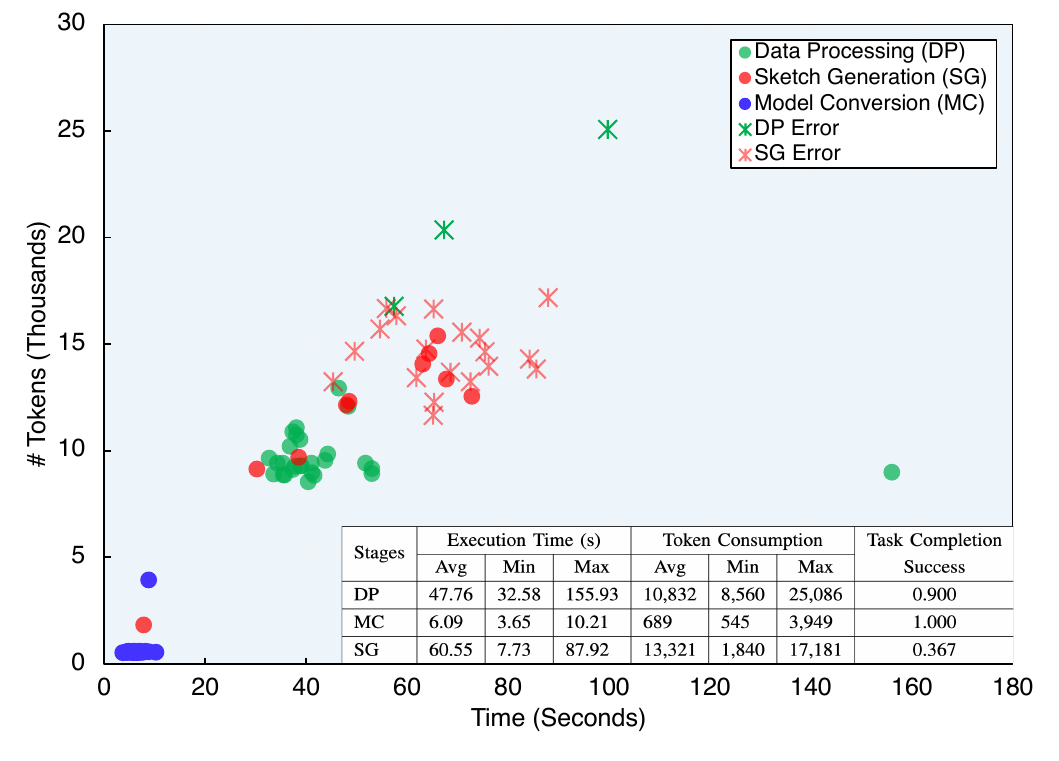}
    \caption{The scatter plot illustrates the relationship between time consumption (in seconds) and token consumption across three TinyML stages: DP, MC, and SG. Successful and failed operations are distinguished by marker types. The accompanying table summarizes average, minimum, and maximum execution times and token consumption for each stage, along with the task completion success rate.}
    \label{fig:scatter}
\end{figure}

\textbf{Empirical Evaluation.} We conducted thirty full runs for each lifecycle stage under evaluation—data processing, 8-bit model quantization and conversion, and sketch generation—using GPT-4o (\texttt{GPT-4o-2024-08-06}). Tokens, the smallest units of text input processed by LLMs, served as a key metric for evaluating resource consumption. To ensure controlled testing, the maximum number of iterations for resolving errors through the framework was capped at \textbf{five} (customizable), preventing excessive retries while allowing prompt adaptation to converge within practical bounds. A stage was considered successful if it completed without LLM failure or compilation error within the retry limit. Execution time was measured from the initial LLM query to the successful completion of the stage. Per-stage results were aggregated over 30 runs, and we reported mean, range, and success rate for each stage to reflect both central tendency and variability. The implementation of the framework spans approximately 1,578 lines of Python code, covering the automation logic, prompt handling, error parsing, and interaction with the LLM API and device toolchain (i.e., \textit{Arduino Command Line Tool} in this case). It is worth highlighting, this end-to-end process introduces no additional computational burden on the target device, as all LLM interactions and code generation occur off-device, and the resulting deployment artifacts are nearly identical to those produced via standard TinyML workflows.  

Among the three TinyML stages, \textit{Sketch Generation (SG)} emerged as the most resource-intensive and least reliable. This stage exhibited a success rate of only 36.7\%, with substantial token consumption averaging 13,321 tokens (range: 1,840–17,181). Execution times were similarly high, averaging 60.55 seconds (range: 7.73–87.92s). In contrast, \textit{Data Processing (DP)} showed moderate resource utilization with an average token consumption of 10,832 tokens (range: 8,560–25,086) and mean execution time of 47.76 seconds (range: 32.58–155.93s), achieving a 90\% success rate. The \textit{Model INT8 Quantization \& Conversion (MC)} stage stood out as the most efficient and reliable, maintaining a 100\% success rate with minimal resource consumption: an average of 689 tokens (range: 545–3,949) and mean execution time of just 6.09 seconds (range: 3.65–10.21s).
  
Fig. \ref{fig:scatter} highlights these performance variations across different TinyML stages, underscoring the specific challenges of the SG phase. The figure also highlights the relationship between execution time and token consumption, revealing distinct and stage-specific patterns. MC clusters tightly in the lower-left quadrant (mean: 6.09s, 689 tokens), demonstrating consistent and efficient performance. This efficiency likely stems from well-defined input/output specifications and standardized TensorFlow Lite procedures. DP, on the other hand, exhibits moderate dispersion. A bimodal distribution is evident, with successful runs consuming fewer resources compared to failed ones, suggesting that failures often result from complex edge cases requiring significantly more computational effort. Sketch Generation presents the most scattered distribution, reflecting substantial variability in resource requirements. Its unpredictability likely arises from the inherent complexity of translating comprehensive specifications into executable C++ code.

The SG stage's high failure rate (63.3\%) and resource consumption variability indicate significant challenges in automated code generation. Failed operations are not simply binary outcomes but involve prolonged computational processes that exhaust resources before declaring failure. This is in contrast to DP and MC, which demonstrate relative stability and efficiency.

In summary, while DP and MC stages are relatively robust and efficient, SG remains a bottleneck due to its lower success rate, higher resource consumption, and variability. These findings highlight the need for further optimization in the automated generation of deployment sketches.

\begin{figure}[!t]
    \centering
    \includegraphics[width=\columnwidth]{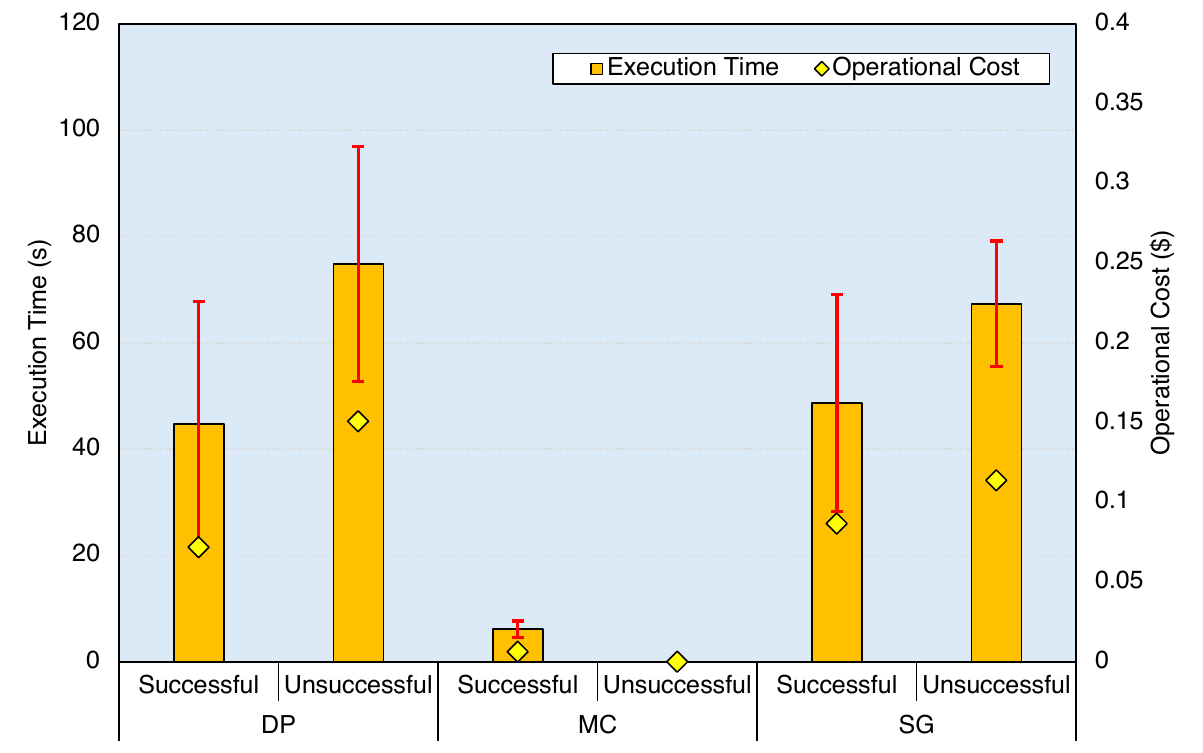}
    \caption{Trade-offs and stage-specific resource consumption concerning execution time (seconds) and operational cost (USD) for successful and unsuccessful runs across DP, MC, and SG stages.}
    \label{fig:costtradeoff}
\end{figure}

To provide deeper insights into the trade-offs across lifecycle stages, we analyzed the relationship between execution time and operational cost for both successful and unsuccessful runs (Fig. \ref{fig:costtradeoff}). We observed that unsuccessful runs consistently exhibit a significant increase in execution time compared to successful ones, highlighting the resource-intensive nature of error-prone processes. This effect is particularly evident in stages like Sketch Generation, where iterative refinements exacerbate execution time variability. In contrast, the operational cost remains relatively stable across both successful and unsuccessful runs, with negligible variance. While this cost may appear minimal at a per-run level, it is critical to consider the cumulative impact in large-scale IoT deployments.

In real-world scenarios involving thousands of devices or frequent lifecycle operations—such as periodic updates for predictive maintenance or environmental monitoring—these costs can accumulate significantly. Moreover, the high variance in execution time introduces unpredictability, which could pose challenges for time-sensitive IoT applications. This analysis underscore the need for more robust and efficient mechanisms, particularly for error-prone stages, to ensure scalability and cost-effectiveness in IoT-centric TinyML workflows.

\textbf{Demo.} To illustrate the practical application of the proposed framework, we provide two video demonstrations. The first video\footnote{\url{https://youtu.be/KnJ5m78x_X8}} highlights the automation of the model quantization and conversion stage, where ML models are optimized for resource-constrained devices and converted into a TensorFlow Lite-compatible format. The second video\footnote{\url{https://youtu.be/Ojpsb5Wnnl8}} demonstrates the automated generation of C++ sketch code for the targeted Arduino, enabling the execution of the converted model.

\section{Challenges and Horizons}

Our work highlights the potential of LLMs to enhance the TinyML lifecycle while uncovering key challenges that must be addressed. This section explores avenues for further development identified during our research.

\textbf{Enhancing Error Resilience.} One significant challenge lies in the system's limited ability to self-correct, especially in multi-step processes where errors in early stages propagate and compound. A promising direction involves considering LLM-based multi-agent systems, combining smaller, locally deployed LLMs with more capable, cloud-based models. Local LLMs could handle error detection and validation, ensuring outputs adhere to expected formats before engaging larger models for more complex tasks. Although performance–cost–bandwidth trade-offs should be evaluated on a case-by-case basis, this approach can help balance privacy, efficiency, and cost, as local models reduce reliance on costly cloud resources. However, computational constraints on TinyML devices make deploying sophisticated models locally challenging, necessitating further optimization and adaptation. In such cases, an alternative strategy could involve equipping the framework, which effectively stands between the end device and the cloud, with a lightweight, embedded LLM acting as a local proxy. This would enable more responsive coordination and selective filtering, improving latency and efficiency in distributed inference workflows. In addition, the use of adaptive prompt templates, where prompt structure evolves dynamically based on execution outcomes and feedback, may further enhance the system’s ability to self-correct. Such templates can prevent repeated failures by refining LLM instructions contextually, improving reliability across iterations.

\textbf{Streamlining Resource-Intensive Processes.} Sketch generation emerged as a bottleneck due to high resource consumption and lower success rates. Enhancing prompt engineering and workflow design could address this, focusing on precise, context-aware prompts and iterative feedback loops to reduce errors. As mentioned above, adaptive prompt templates represent a promising technique in this direction, enabling more reliable and tailored code generation through dynamic prompt evolution. Additionally, exploring domain-specific languages (DSLs) tailored for target platforms, such as Arduino, could streamline code generation. As an example, DSLs inspired by symbolic programming languages, such as Prolog, have been investigated for improving reasoning and control logic in constrained environments, offering an avenue to complement LLM-based generation with more structured and explainable constructs. Finally, hybrid approaches that combine LLM-generated drafts with traditional compilers and manual developer refinement may further improve reliability and efficiency, especially for complex or resource-intensive tasks.

\textbf{Adaptive Task Allocation for Cost Efficiency.} Monetary costs associated with API calls become significant at scale, particularly in dynamic IoT deployments. Adaptive LLM routing offers a solution, directing simpler tasks to smaller, cost-effective models while reserving more complex queries for larger models. For instance, tasks such as data processing or model conversion—which in our evaluation showed higher success rates—can be routed to LLMs with lower-cost APIs. This introduces a trade-off: slightly lower reliability may require more iterations, but can yield meaningful cost savings. This strategy can extend to specialized LLMs for TinyML, enabling dynamic updates or functionality shifts, such as adding new features or reconfiguring device applications. For example, if an application initially performs object detection but later requires object tracking, the framework can route the request to a model more specialized in temporal or visual sequence tasks. By acting as a capability discovery system, the framework can negotiate with specialized LLMs to identify the most efficient solution, optimizing cost and performance while maintaining adaptability for evolving TinyML requirements.

\section{Conclusions: Reality, Illusion, or Opportunity?}

In this paper, we explored the integration of LLMs into the TinyML lifecycle, investigating their potential to automate and streamline key stages through the development of a lifecycle middleware framework tested through a practical IoT development scenario involving a classification use case. Our findings reveal both the promise and challenges of this approach. 
On the positive side, LLMs demonstrate the ability to improve development efficiency, as seen in tasks like model quantization and data preprocessing. However, limitations remain: reliability issues in complex tasks such as sketch generation, constrained generalizability across hardware platforms, and significant resource demands in terms of computation and cost. These challenges highlight the need for further refinement, including enhanced prompt engineering, specialized model fine-tuning, and the development of advanced symbolic reasoning mechanisms to improve error correction, iterative refinement, and reasoning capabilities.
Despite these hurdles, the \textbf{opportunity} for transformation is clear. By addressing these challenges and expanding the framework to include additional lifecycle stages and more diverse hardware, LLMs can become a cornerstone of TinyML automation. This work lays the groundwork for leveraging LLMs alongside traditional tools, paving the way for more efficient, scalable, and accessible embedded IoT ML workflows. Through the development of this end-to-end framework for automating core TinyML lifecycle stages using LLMs, we contribute a novel methodology that reduces developer effort and highlights the feasibility of language model–driven embedded ML workflows. Looking forward, open questions remain about the long-term generalizability of LLM-generated code across highly heterogeneous hardware platforms, and how such models can reliably adapt to evolving application needs with minimal supervision. Addressing these questions may define the next phase of LLM-enabled intelligence at the edge.

\section{Acknowledgements}
This work was supported by the \textit{Digital Twinning of Personal Area Networks for Optimized Sensing and Communication} project through the Business Finland 6G Bridge Program (8782/31/2022).

\bibliographystyle{IEEEtran}
\bibliography{References}
\vskip -2\baselineskip plus -1fil
\begin{IEEEbiographynophoto}{Guanghan Wu}(Student Member, IEEE) is a Research Assistant in the Department of Computer Science at the University of Helsinki, Finland. He earned his Master’s degree in Computer Science from the same university in 2024. (guanghan.wu@helsinki.fi)
\end{IEEEbiographynophoto}
\vskip -2\baselineskip plus -1fil
\begin{IEEEbiographynophoto}{Sasu Tarkoma}(Senior Member, IEEE)  is a Professor of Computer Science and Dean of the Faculty of Science at the University of Helsinki, Finland. He is affiliated with the Helsinki Institute for Information Technology (HIIT) and the Finnish Center for AI (FCAI). He is the chairman of the Finnish Scientific Advisory Board for Defence (MATINE). His research interests include Internet and distributed systems, IoT, mobile computing, and AI. (sasu.tarkoma@helsinki.fi)
\end{IEEEbiographynophoto}
\vskip -2\baselineskip plus -1fil
\begin{IEEEbiographynophoto}{Roberto Morabito} (Member, IEEE) is an Assistant Professor in the Communication Systems Department at EURECOM, France. His research focuses on networked AI systems, with particular attention to AI service provisioning and lifecycle management under computing and networking constraints. He earned his PhD from Aalto University and has held positions at the University of Helsinki, Princeton University, and Ericsson Research Finland. (roberto.morabito@eurecom.fr)
\end{IEEEbiographynophoto}

\end{document}